\documentclass[aps,preprintnumbers,notitlepage,tightenlines, groupedaddress,nofootinbib, amsmath,amsfonts,amssymb]{revtex4-1}

\usepackage{hyperref}
\usepackage{graphicx}
\usepackage{braket}

\usepackage{color}

\newcommand{\drm}{\mathrm{d}}
\newcommand{\der}[2]{\frac{\drm #1}{\drm #2}}
\newcommand{\vev}[1]{\langle {#1}\rangle}
\newcommand{\dg}{\dagger}
\DeclareMathOperator{\tr}{\mathrm{tr}}

\newcommand{\be}{\begin{equation}}
\newcommand{\ee}{\end{equation}}
\newcommand{\bea}{\begin{eqnarray}}
\newcommand{\eea}{\end{eqnarray}}
\newcommand{\ud}{\mathrm{d}}

\newcommand{\reef}[1]{\eqref{#1}}

\begin{document}

\preprint{Brown - HET-1657}

\title{Universal Bounds on the Time Evolution of Entanglement Entropy}

\author{Steven G. Avery}
\email{steven\textunderscore{}avery@brown.edu}
\author{Miguel F. Paulos}
\email{miguel\textunderscore{}paulos@brown.edu}
\affiliation{Department of Physics,
Brown University,
Box 1843,
Providence, RI 02912-1843,
USA}

\date{\today}

\begin{abstract}
Using relative entropy, we derive bounds on the time rate of change of geometric entanglement entropy for any relativistic quantum field theory in any dimension. The bounds apply to both mixed and pure states, and may be extended to curved space. We illustrate the bounds in a few examples and comment on potential applications and future extensions.
\end{abstract}


\maketitle

\section{Motivation and Introduction}

Recently, entanglement entropy has become an important theoretical
tool for probing quantum physics in diverse situations.  Of especial
interest is the geometric entanglement entropy (GEE), $S_V$,
associated with some spatial region, $V$. To wit, the von Neumann
entropy of the reduced density matrix found by tracing out the degrees
of freedom associated with the complementary region,
$\bar{V}$.\footnote{We do not worry about issues associated with the
  ability to decompose the Hilbert space into a tensor product,
  $\mathcal{H} = \mathcal{H}_V\otimes
  \mathcal{H}_{\bar{V}}$. See~\cite{Casini:2013rba}, for a recent
  discussion.}

In this paper, we are interested in how causality and locality bound
the rate of change of entanglement entropy, $\der{S}{t}$, for excited
states in a relativistic quantum field theory. As is now well-known,
the geometric entanglement entropy is UV divergent in the vacuum, with
the leading divergence proportional to the area of $\partial
V$. Because of this UV sensitivity, one might question whether there
are any interesting bounds at all; however, entropy differences are
frequently finite for reasonable states, and therefore one should
expect that $\der{S}{t}$ is UV finite for reasonable states. This is
also supported by some previous explicit calculations,
cf.~\cite{Calabrese:2005in,Calabrese:2009qy,Liu:2013qca,Liu:2013iza}.

There are two relevant bodies of research in the literature. Firstly,
there are bounds on $\der{S}{t}$ for finite-dimensional
nonrelativistic quantum mechanical systems. The most relevant to us is
the proof of the small incremental entangling (SIE) conjecture
in~\cite{PhysRevLett.111.170501}, building on work
in~\cite{PhysRevA.76.052319}. The SIE conjecture states that for a
four-part system $aABb$ evolving with Hamiltonian of the form $H =
H_{aA} + H_{bB} + H_{AB}$,\footnote{Systems $a$ and $b$ are called
  ancilla, since they do not directly interact with each other.} the
maximum growth of the entanglement entropy of $aA$ is
bounded~\cite{PhysRevA.76.052319}:
\begin{equation}\label{eq:anc-bound}
\der{S_{aA}}{t}\bigg|_\text{max} \leq k\, \|H_{AB}\|\log d\qquad d=\min(d_A, d_B),
\end{equation}
where $k$ is an order unity constant, $\|H_{AB}\|$ is the operator
norm of the interacting Hamiltonian, and the maximum is taken over all
states. This can be used to argue that if one state obeys the area
law, then all adiabatically connected states do as
well~\cite{PhysRevA.76.052319,PhysRevLett.111.170501}; for a lattice
system $\log d\simeq A\,\log d_s$, where $d_s$ is the dimension of
each lattice site's Hilbert space and $A$ is the area measured in
lattice units.

Unfortunately, it is difficult to directly apply this to quantum field
theory, since even in lattice QFT the per-site Hilbert space is
infinite dimensional.\footnote{Although see~\cite{Jordan:2011ne} for a
  regularization scheme that uses a finite dimensional Hilbert space.}
Moreover, Lorentz invariance would enter into this argument only
indirectly in the form of the Hamiltonian. Finally, let us note that
there are states for which $\der{S}{t}$ \emph{is} UV
divergent~\cite{Calabrese:2007,Asplund:2011cq,Asplund:2013zba}, and
thus we expect that this probably is not even the right starting
point. These examples have divergent stress tensors $T_{\mu\nu}$, and
since the accessible phase space grows with energy scale, it is
perhaps not surprising that $\der{S}{t}$ diverges.

The divergent dimension of the Hilbert space and the existence of
states with diverging $\der{S}{t}$ are actually related issues. The
above bound is derived after maximizing over the \emph{entire} Hilbert
space. This is a sensible thing to do in quantum mechanics, but is not
consistent with our modern Wilsonian understanding of QFT since this
maximization would be over states with arbitrarily large energies and
momenta. If we allow ourselves to discuss additional information about
the scale of the state, then we can hope that there is some
finite-dimensional subspace and a bound may
exist. (See~\cite{Swingle:2013hga}, for example.) For our discussion,
we are content to suppose we know $\vev{T_{\mu\nu}(x)}$, which allows
us to connect entanglement and energy/momentum. (In fact, we write our
bound in terms of the expectation value of the modular Hamiltonian,
which is a linear functional of the stress tensor in the few cases for
which we have an explicit, local expression.)

The second body of literature concerns fundamental relativistic bounds on the
transmission rate of classical
information---\cite{Bekenstein:1990du} gives an extensive
discussion. In so far as the von Neumann entropy is the quantum
analogue of the classical Shannon entropy, and that many bounds on
classical information carry over to analogous bounds on quantum
information~\cite{nielsen:2000}, it is natural to ask whether these
bounds have quantum anologues as bounds on $\der{S}{t}$. The most
important bound for us derives from the Bekenstein
bound~\cite{Bekenstein:1980jp}:
\begin{equation}
H \leq \frac{2\pi E R}{\hbar c},
\end{equation}
where $H$ is the thermodynamic entropy, $E$ the energy of some object
that can be circumscribed by a radius $R$ ball. While this bound
originated from black hole thermodynamics, it is supposed to be valid
for any system one can throw into a black hole. If one considers
information transmission via material transport, then one finds
that~\cite{Bekenstein:1981zz}
\begin{equation}\label{eq:classical-inf-bound}
\dot{I} \leq \frac{2\pi E}{\hbar},
\end{equation}
where $\dot{I}$ is the classical communication rate measured in
``nats'' per unit time.\footnote{A nat is $(\ln 2)^{-1}\approx 1.44$ bits.}

Unfortunately, the precise range of applicability and validity of the
Bekenstein bound is obscured by ambiguities in defining all three
related quantities: $H$, $E$, and $R$. The original argument for the
bound has also been challenged~\cite{Unruh:1982ic,Unruh:1983ir};
see~\cite{Bousso:2003cm, Bekenstein:2004sh} for recent defenses of the
bound.

Fortunately, positivity of relative entropy provides an apodictic
quantum analogue of the Bekenstein bound, which is not plagued by the
same
ambiguities~\cite{Blanco:2013joa,Casini:2008cr}.\footnote{In~\cite{Blanco:2013lea}
  another related inequality, which we do not find useful here, was
  called a ``Bekenstein bound''.} Since the original Bekenstein bound
immediately led to a bound on the transmission of classical
information, one should guess that the new refined version should
imply a bound on $\der{S}{t}$. In fact, the calculation is not as
straightforward as the classical case, because we must carefully
formulate bounds that subtract off contributions from the
vacuum. Instead of using the positivity of relative entropy, we
primarily use the monotonicity property.

\section{Derivation}

\subsection{Causal Domains}\label{sec:causal}

We begin our derivation by first noting that we are working with a
relativistic QFT in $d$-dimensional Minkowski space. We are interested in the
entanglement entropy of a region $V$ as a function of $t$. Let us consider
evaluating $\der{S}{t}$ as usual in a limiting procedure via
\begin{equation}
\der{S_V(t)}{t} = \lim_{\delta t\to 0} \frac{S_V(t + \delta t) - S_V(t)}{\delta t}.
\end{equation}
Since entropy differences are finite for reasonable states, we expect
this to be finite for ``nice'' states. We spend most of our effort
manipulating the entropy difference in the numerator.

Let $\Sigma_t$ denote the spatial slice at time $t$, so that $V\subset
\Sigma_t$. In this language, $V'\subset \Sigma_{t+\delta t}$ is the
time translation of $V$ and $\der{S}{t} = \lim_{\delta t\to 0} (S_{V'}-S_{V})/\delta t$ .
The causal domain of $V$, $\mathcal{D}(V) = \mathcal{D}^+(V)\cup
\mathcal{D}^-(V)$, is given by the set of events for which either the
past or future lightcone intersects $\Sigma_t$ as a subset of $V$. The
GEE $S_V$ more correctly is a function of $\mathcal{D}(V)$, since changes in
the slicing $\Sigma_t$ that keep $\partial V$ fixed effect unitary
transformations on the density matrix and leave $S_V$
invariant. 

\begin{figure}[t]
\includegraphics[height=6cm]{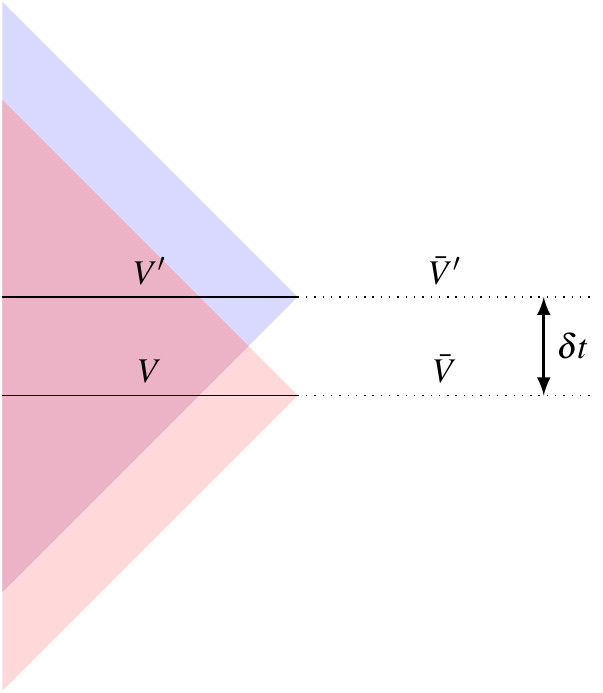}
\hspace*{1cm}
\includegraphics[height=6cm]{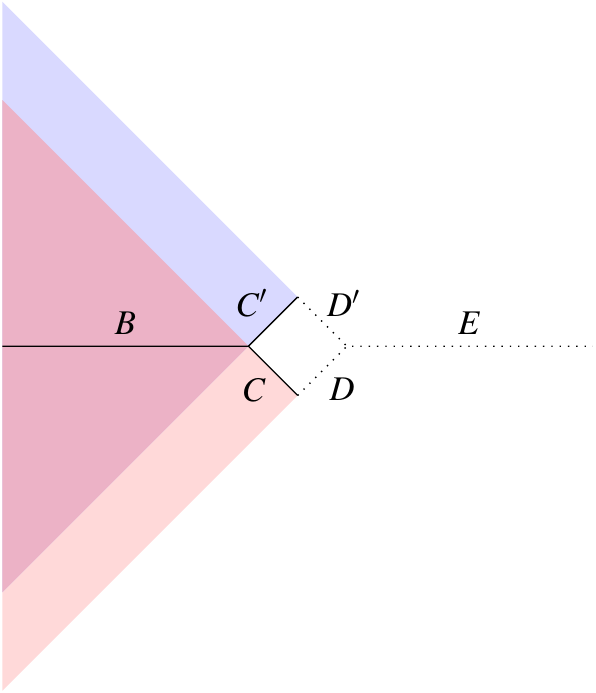}
\caption{\label{fig:causal} Entanglement entropy is invariant under
  different spatial slicings that preserve the causal domain. Thus, we
  may deform the evolution from $S_V$ to $S_{V'}$ into evolution from
  $S_{BC}$ to $SC_{BC'}$ as shown above for the half space. This
  isolates all of the interesting dynamics into a $\delta t$-size
  ``diamond''.}
\end{figure}

Thus, we can deform the spatial slices inside $\partial V$ or outside
$\partial V$ at the two times without changing the answer. It seems
convenient to deform the two slices as shown in
Figure~\ref{fig:causal}. We decompose the slices into an invariant
spatial region $B$, followed by two (in the limit) null regions $C$
and $D$, and another invariant spatial region $E$. The total state on
$BCDE$ and $BC'D'E$ will be pure if the total system is in a pure
state. This slightly singular evolution\footnote{See the recent
  paper~\cite{Bousso:2014uxa}, for some interesting results and
  subtleties related to null surfaces.} gives us two states related by
a unitary transformation that acts only on the the $CD$ space:
\begin{equation}
\ket{\psi_{BC'D'E}} = U_{CD} \ket{\psi_{BCDE}}.
\end{equation}
The original density matrices for $V$, $\rho_V$ and $\rho_{V'}$, are related by unitary
transformations to
\begin{equation}
\rho_V = U_1\rho_{BC}U_1^\dg\qquad
\rho_{V'} = U_2\rho_{BC'}U_2^\dg.
\end{equation}
Hence, the entropy is the same.  Note that the transformation from
$\rho_{BC}$ to $\rho_{BC'}$ looks like a quantum operation that
depends on the state of $C$. The regions $B$ and $E$ seem to play the
role of ancilla, although keep in mind that we are going to be taking
the limit as $\delta t\to 0$ and these regions all depend on $\delta t$.

Formally, we can define the above regions as follows:
\begin{equation}
\begin{gathered}
C  = \partial \mathcal{D}^+(V) \cap \overline{\mathcal{D}^-(V')} \qquad
C' = \partial \mathcal{D}^-(V')\cap \overline{\mathcal{D}^+(V)}\\
D  = \partial \mathcal{D}^+(\bar{V})\cap \overline{\mathcal{D}^-(\bar{V}')} \qquad
D' = \partial \mathcal{D}^-(\bar{V}')\cap \overline{\mathcal{D}^+(\bar{V})}\\
B  \subset \mathcal{D}^-(V')\cap \mathcal{D}^+(V)\text{ with } 
    \partial B = \partial \mathcal{D}^-(V')\cap \partial \mathcal{D}^+(V).
\end{gathered}
\end{equation}
The various regions are illustrated in Figure~\ref{fig:causal} for the
half space.  We now need to bound $S_{BC'}-S_{BC}$ in the limit of
small $\delta t$. Note these definitions suggest a clear
generalization to curved background metrics.
\subsection{Relative Entropy}

Recently, it was pointed out that the relative entropy furnishes a
more precise version of the Bekenstein bound \cite{Casini:2008cr}. Recall that the relative
entropy is a measure of the distinguishability of a density matrix
$\rho$ from a density matrix $\sigma$ given by\footnote{We use
  $S(\cdot\|\cdot)$ instead of $S(\cdot|\cdot)$ to distinguish the
  relative entropy from the conditional entropy.}~\cite{nielsen:2000}
\begin{equation}
S(\rho\|\sigma) = \tr \rho\log \rho -\tr \rho\log\sigma.
\end{equation}
Note the asymmetry between $\rho$ and $\sigma$.

The relative entropy satisfies two inequalities~\cite{nielsen:2000}
that are important for our purposes. First, Klein's inequality:
$S(\rho\|\sigma) \geq 0$ with equality if and only if
$\rho=\sigma$. Second, the relative entropy monotonically decreases
under partial tracing: $S(\rho_{\alpha\beta}\|\sigma_{\alpha\beta})
\geq S(\rho_\alpha\|\sigma_\alpha)$. Heuristically, decreasing the
number of degrees of freedom one can access decreases
distinguishability.

As noted in~\cite{Casini:2008cr}, if we write $\sigma = N
e^{-K}$,\footnote{The normalization $N$ can be fixed by demanding $\sigma$
  have unit trace.} then the relative entropy can be cleverly
rewritten as
\begin{equation}
S(\rho\|\sigma) = \Delta\vev{K} - \Delta S,
\end{equation}
where $\Delta$ indicates the difference of the quantity when evaluated
in state $\rho$ from state $\sigma$. We will always take $\sigma$ to
be the reduced density matrix one gets from the vacuum, for which $K$
is the modular Hamiltonian. Then the nonnegativity of the relative
entropy implies an upper bound on the regulated (vacuum-subtracted)
entropy $\Delta S$,
\begin{equation}
\Delta S \leq \Delta\vev{K}.
\end{equation}
In the cases where we understand the modular Hamiltonian $K$, this
bears a remarkable similarity to the original Bekenstein
bound~\cite{Casini:2008cr}.

\subsection{Bounds}

We can now use the monotonicity property of relative entropy for the
regions defined in Section~\ref{sec:causal}. First note that
monotonicity implies
\begin{equation}
S(\rho_{BCD}\|\sigma_{BCD})\geq S(\rho_B\|\sigma_B)\quad \Longrightarrow\quad
\Delta S_{BCD}-\Delta S_{B} \leq \Delta\vev{K_{BCD}} - \Delta\vev{K_B}.
\end{equation}
We also have an equivalent bound for the complementary regions:
\begin{equation}
\Delta\vev{K_E}-\Delta \vev{K_{CDE}} \leq \Delta S_E - \Delta S_{CDE}
\end{equation}
We want to relate the LHS of the first inequality to the RHS of
the second. If the total state of the QFT is pure, then the
two quantities are equal since $S=\bar{S}$ for a pure state; but if the total state is mixed, for instance thermal, then we have to
work a little harder. First note that strong subadditivity (SSA) of
entanglement implies
\begin{equation}
S_E - S_{CDE}\leq S_{BCD} - S_B.
\end{equation}
Unfortunately, SSA does not directly apply to the regulated
entanglement entropy. In this case, however, purity of the vacuum
implies
\begin{equation}
 S^\text{vac}_E - S^\text{vac}_{CDE} = S^\text{vac}_{BCD}-S^\text{vac}_B,
\end{equation}
and therefore
\begin{equation}
\Delta S_E -\Delta S_{CDE} \leq \Delta S_{BCD}-\Delta S_B.
\end{equation}
This allows us to write
\begin{equation}\label{eq:delta-x-bound}
\Delta\vev{K_E}-\Delta \vev{K_{CDE}} \leq \Delta S_E - \Delta S_{CDE}
\leq \Delta S_{BCD}-\Delta S_{B} \leq \Delta\vev{K_{BCD}} - \Delta\vev{K_B}
\end{equation}
Interestingly, this inequality holds as long as either $\rho$ or
$\sigma$ come from a pure state. Dividing by $\delta t$ and taking
$\delta t$ to zero, this becomes an upper and lower bound on the
normal derivative of the regulated entanglement entropy in terms of
normal derivatives of modular hamiltonians.

Monotonicity implies
\begin{equation}
S(\rho_B\|\sigma_B)\leq S(\rho_{BC}\|\sigma_{BC})\leq S(\rho_{BCD}\|\sigma_{BCD})
\end{equation}
together with
\begin{equation}
-S(\rho_{BCD}\|\sigma_{BCD}) \leq -S(\rho_{BC'}\|\sigma_{BC'}) \leq -S(\rho_B\|\sigma_B),
\end{equation}
as well as the equivalent relations for the complementary
region. Judicious use of the inequalities allows us to write
\begin{align}
S(\rho_{BC'}\|\sigma_{BC'}) - S(\rho_{BC}\|\sigma_{BC}) &\geq S(\rho_{B}\|\sigma_{B})
    -S(\rho_{CDE}\|\sigma_{CDE})-\Delta\vev{K_{BC}}+\Delta\vev{K_{DE}}\\
S(\rho_{BC'}\|\sigma_{BC'}) - S(\rho_{BC}\|\sigma_{BC}) &\geq S(\rho_{E}\|\sigma_{E})
    -S(\rho_{BCD}\|\sigma_{BCD})+\Delta\vev{K_{BC'}}-\Delta\vev{K_{D'E}}.
\end{align}
While the two inequalities hold separately, it is convenient to add them to find
\begin{multline}\label{eq:delta-upper}
2[S(\rho_{BC'}\|\sigma_{BC'}) - S(\rho_{BC}\|\sigma_{BC})] \geq\\
\Delta\vev{K_{BC'}}-\Delta\vev{K_{D'E}}-\Delta\vev{K_{BC}}+\Delta\vev{K_{DE}}
+\Delta\vev{K_{B}}-\Delta\vev{K_{CDE}}\\
-\Delta\vev{K_{BCD}}+\Delta\vev{K_{E}}
+[\Delta S_{BCD}-\Delta S_E+\Delta S_{CDE}-\Delta S_B].
\end{multline}
The last term in brackets we can drop since it is positive definite
from~\eqref{eq:delta-x-bound}. Using the same techniques, we can find
an upper bound as well:
\begin{multline}\label{eq:delta-lower}
2[S(\rho_{BC'}\|\sigma_{BC'}) - S(\rho_{BC}\|\sigma_{BC})] \leq\\
\Delta\vev{K_{BC'}}-\Delta\vev{K_{D'E}}-\Delta\vev{K_{BC}}+\Delta\vev{K_{DE}}
-\Delta\vev{K_{B}}+\Delta\vev{K_{CDE}}\\
+\Delta\vev{K_{BCD}}-\Delta\vev{K_{E}}.
\end{multline}


Let us define the time and normal derivatives as
\begin{subequations}\label{eq:derivatives}
\begin{align}
\der{S}{t} &= \der{\Delta S}{t} = \lim_{\delta t \to 0}\frac{S_{BC'}-S_{BC}}{\delta t}\\
\der{\Delta S}{x_\perp} &= \lim_{\delta t\to 0} \frac{\Delta S_{BCD} - \Delta S_B}{\delta t}\\
\der{\Delta\bar{S}}{x_\perp} &= -\lim_{\delta t\to 0}
     \frac{\Delta S_{CDE}-\Delta S_{E}}{\delta t},
\end{align}
\end{subequations}
with the obvious parallel definitions for $K$s. Note that time
translation invariance of the vacuum means that the vacuum subtraction
drops out from the time derivatives; this is \emph{not} true for the
normal derivative since the vacuum entanglement is not invariant under
increases in the region size. Also, note the $x_\perp$ is oriented
outward from $V$.

The normal derivative defined in~\eqref{eq:derivatives} deserves some
explication. For an arbitrary region at constant $t$ with a smooth boundary $\ud
x_\perp$ is a normal shift of the boundary. To be precise, if we
consider some $F$ which is a functional of the entangling surface
parametrized by $x^{\mu}(s_a)$, with $\mu=0,\ldots, d-1$, $a=1,\ldots d-2$
then we have
\begin{equation}
\der{F}{x_\perp}\equiv \int \ud^{d-2} s \frac{\delta F}{\delta x^{\mu}(s)} \frac{\omega^{\mu \nu} \hat{t}_\nu}{|\omega|},\qquad 
\omega_{\mu \nu}=\epsilon_{\mu \nu \rho_1\ldots \rho_{d-2}} \frac{\partial x^{\rho_1}}{\partial s^1}\ldots \frac{\partial x^{\rho_{d-2}}}{\partial s^{d-2}},
\end{equation}
with $\hat{t}^\mu$ being the unit time vector. With this notation,
Equation~\eqref{eq:delta-x-bound} becomes
\begin{equation}\label{eq:der-x-bound}
\der{\Delta\vev{\bar{K}}}{x_\perp}\leq \der{\Delta \bar{S}}{x_\perp}\leq \der{\Delta S}{x_\perp}
   \leq \der{\Delta \vev{K}}{x_\perp},
\end{equation}
and the two inequalities~\eqref{eq:delta-upper} and~\eqref{eq:delta-lower} become
upper and lower bounds on $\der{S}{t}$:
\begin{equation}\begin{aligned}
\der{S}{t} &\geq \frac{1}{2}\der{}{t}\bigg(\Delta\vev{K}+\Delta\vev{\bar{K}}\bigg)
  - \frac{1}{2}\der{}{x_\perp}\bigg(\Delta\vev{K}-\Delta\vev{\bar{K}}\bigg)\\
\der{S}{t} &\leq
\frac{1}{2}\der{}{t}\bigg(\Delta\vev{K}+\Delta\vev{\bar{K}}\bigg) 
   + \frac{1}{2}\der{}{x_\perp}\bigg(\Delta\vev{K}-\Delta\vev{\bar{K}}\bigg),
\end{aligned}\end{equation}
where note that $\der{(K-\bar{K})}{x_\perp}$ is nonnegative
from~\eqref{eq:der-x-bound}. In fact, there are many other bounds one
can derive from monotonicity; however, in practice these seem to be
the tightest bounds on $\der{S}{t}$. These bounds hold universally for
any unitary Lorentz invariant theory, but for the cases in which the
modular Hamiltonian is local and known, we may write them directly in
terms of the stress tensor and simplify them further. We shall now
consider the two better known cases in turn.


\section{Illustration}

\subsection{The Half-Space}

In our first example the region $V$ is the half-space $x_1<X$. It is a
quite general result~\cite{Bisognano1975,Bisognano1976} that for this geometry the modular hamiltonian
corresponding to the vacuum state necessarily becomes the boost charge. That is,
\begin{subequations}\begin{align}
\Delta K &= 2\pi \int_{x_1<X} \ud^{d-1}x\, (X-x_1)\, T^{00}(t,\vec x), \\
\Delta \bar K &= 2\pi \int_{x_1>X} \ud^{d-1}x\, (x_1-X)\, T^{00}(t,\vec x).
\end{align}\end{subequations}
A simple computation yields
\begin{subequations}\begin{align}
\der{\vev{K}}{t}&= -2\pi \int_{x_1>0} T^{01}(t,\vec x)=-2\pi P^1\qquad
 &\der{\vev{\bar{K}}}{t} &= 2\pi \bar{P}^1 \\
\der{\Delta\vev{K}}{X}&= 2\pi P^0\qquad &\der{\Delta\vev{\bar K}}{X}&= -2\pi \bar P^0
\end{align}\end{subequations}
with $P^1$ the momentum along $x_1$ of the half-space, $P^0$ its
energy, along with $\bar P^0$ and $\bar{P}^1$ the energy and momentum
of its complement. Defining the total energy $E_T=P^0+\bar P^0$ and
total momentum $P^1_T=P^1 + \bar{P}^1$, the bounds become
\begin{equation}
-2\pi P^1 + \pi(P^1_T - E_T)\leq \der{S}{t} \leq -2\pi P^1 + \pi (P^1_T + E_T) \label{boundhalfspace}
\end{equation}
Note the qualitative similarity with the classical
bound~\eqref{eq:classical-inf-bound}.

\subsection{The Ball}
Now consider the case where the region $V$ is the ball of radius $R$
centered at the origin. If we are dealing with a \emph{conformal} field
theory, we may use a conformal mapping from the Rindler wedge onto the
causal development of the ball to obtain the modular hamiltonian~\cite{Hislop1982,Casini2011},
\bea
\Delta K&=& \pi \left(R\, P^0-K^0/R\right)=\pi \int_{r<R} \ud^{d-1}x\, \frac{R^2-r^2}{R}\, T^{00}(t,\vec x) \\
\Delta \bar K&=& - \pi \left(R\, \bar P^0-\bar K^0/R\right)=\pi \int_{r>R} \ud^{d-1}x\, \frac{r^2-R^2}{R}\, T^{00}(t,\vec x).
\eea
Notice that for $r\simeq R$ we recover the result of the previous
section.  As before we can use conservation of the
stress-tensor to obtain
\begin{subequations}\begin{align}
\der{\vev{K}}{t}&= - \frac{2\pi}R\,  \int_{r<R} \ud^{d-1}x\, x^i T^{i0}=- \frac{2\pi}{R} D, \qquad 
&\der{\vev{\bar K}}{t}&=\frac{2\pi}{R} \bar D
\\
\der{\Delta\vev{K}}{R}&=\pi P^0+\pi K^0/R^2, \qquad &\der{\Delta\vev{K}}{R}&=-\pi \bar P^0-\pi \bar K^0/R^2,
\end{align}
\end{subequations}
with $D$ the dilatation charge. In this way the bounds become
%
%
\begin{equation}
-\frac{2\pi}{R} D + \frac{\pi}{2}\left(\frac{2D_T}{R}- E_T -\frac{K^0_T}{R}\right) \leq\der{S}{t}
\leq -\frac{2\pi}{R} D + \frac{\pi}{2}\left(\frac{2D_T}{R}+E_T+\frac{K^0_T}{R^2}\right)
\end{equation}
We can imagine increasing the radius of the sphere and simultaneously translating it to obtain the half-space. In this limit we have
\bea
D\to R\,P^1, \qquad K^0_T\to R^2 E_T,
\eea
and we recover the bounds \reef{boundhalfspace}.

\section{Conclusion}

We have derived bounds on the entangling rate valid for any unitary
Lorentz invariant quantum field theory in any dimension. We shall not
show it here, but we have checked they are satisfied in all cases
where we were able to easily test them.  The bound can be thought of
as a quantum version of the structurally similar bound on
classical information in~\eqref{eq:classical-inf-bound}. One way of
thinking about this is in terms of mutual information: if the whole
system $AB$ is in a pure state, then the mutual information between
region $A$ and region $B$ reduces to $2 S_A$---so our bounds
impose constraints on the rate of information flow from one region to
the other. We find that just as for the classical case, there seems to
be an energy cost associated to the transmission of quantum
information. Much like what happens for the proposed quantum version
of the Bekenstein bound involving relative
entropy~\cite{Casini:2008cr}, our bounds resolve many of the
ambiguities inherent in equation~\eqref{eq:classical-inf-bound} by
using quantities measured relative to the vacuum state.

As it stands, our bounds hold even for theories without a local
stress-tensor---such as defect or boundary CFTs. This goes some way to
explaining why our bounds involve global charges even when the modular
hamiltonian has a local expression. But it also suggests these bounds
can be made stronger. For instance, considering a distant perturbation
that increases the total energy, we expect that $\der{S}{t}$ should
vanish until signals from the perturbing event could possibly reach the
region, at least for local field theories. Unfortunately, our bound
does not seem to account for this aspect of causality, since
independently of distance these perturbations still affect global
charges such as total energy.

One idea for improving the bounds is to focus our attention on the
dynamics inside the small causal diamond at the boundary of the causal
developments of two Cauchy slices separated by a small $\delta t$,
such as the diamond bounded by regions $C$ ,$C'$ ,$D$, and $D'$ in
Figure~\ref{fig:causal}. In this small region, we are probing the UV
dynamics of the theory. Assuming a free UV fixed point, then it seems
that to leading order the leading process can only be a ``swap gate'':
$D\to C'$ and $C\to D'$. Using this should be enough to derive a
stronger, local bound. And yet, we should offer a word of caution:
simple dimensional analysis seems to preclude a linear and
local bound, at least for the half-space geometry, unless we are
willing to introduce some cutoff dependence; although something
non-linear such as
\begin{equation}
\left |\frac{\ud S}{\ud t}\right|^2\leq \frac{\ud^2 K}{\ud t\, \ud x} \simeq \int_{\partial V}T^{00}.
\end{equation}
is perfectly fine. In fact a bound on classical information very
similar to the above appears in the literature;
see~\cite{Bekenstein:1990du}, and references therein. Let us also
note, one may derive a bound that maximizes over the Hilbert space
like that in~\eqref{eq:anc-bound} by using Bousso's covariant entropy
bound~\cite{Bousso:2002ju} as a cutoff; however, such a bound, being
in Planck units, would have limited utility.

Another possibility is to consider bounds on the second time
derivative. If we consider the relative entropy between states at
times $t$ and $t+\delta t$ it is easy to derive
\begin{equation}
\frac{\ud S}{\ud t}=\frac{\ud\langle  K\rangle}{\ud t}, \qquad \frac{\ud^2 S}{\ud t^2}\leq \frac{\ud^2 \langle K\rangle}{\ud t^2}
\end{equation}
Unfortunately, $K$ here is the modular hamiltonian for the system in
the state at time $t$, which is inaccessible in general. We have also
tried an approach in the same lines as those in this paper, by
considering three closely space Cauchy slices, and using monotonicity
of relative entropy. However, and quite generally, we were not able to
find any such bound.

One obvious extension of our work here is to examine what the bounds
imply for holographic entanglement
entropy~\cite{Ryu:2006bv,Ryu:2006ef,Hubeny:2007xt}. The bound, or a
suitable extension of it for curved space, may have implications for
black hole evaporation and the recent black hole entanglement
crisis~\cite{Mathur:2009hf,Braunstein:2009my,Giddings:2011ks,Avery:2011nb,Almheiri:2012rt}. As
in the discussion of~\cite{PhysRevA.76.052319,PhysRevLett.111.170501},
we may also use the bound to tell us about entanglement in the vacuum
of adiabatically connected theories.

\begin{acknowledgments}

This paper benefited from discussions with C.~Asplund, B.~Chowdhury,
M.~Headrick, A.~Maloney, and B.~Schwab. Both authors are supported by
DOE Grant: DE-SC0010010.

\end{acknowledgments}

\bibliography{entBound}
\end{document}